\documentclass{iopart}
\usepackage[dvips]{graphicx}
\usepackage{iopams}
\usepackage{epsf}
\begin{document}
\baselineskip 14pt
\noindent

\begin{tabbing}
\hspace{11cm}\=\kill
 \> \today.
\end{tabbing}

\vspace{1cm}
\noindent
\begin{center}
{\LARGE  An EPMC study of ZnCl$_2$ melt}

\vspace{1cm}
\noindent
{\large J. Neuefeind} \\[.5cm]

\begin{minipage}[c]{10cm}
Hamburger Synchrotronstrahlungslabor HASYLAB at 
Deutsches Elektronensynchrotron DESY, 
Notkestr 85, D-22603 Hamburg Germany\\
\end{minipage}\\[.5cm]

\vspace{1cm}
\begin{abstract}
The partial structure factors (PSF) of molten ZnCl$_2$ are known since 1981
due to the neutron diffraction experiments by Biggin and Enderby. It is shown 
in this article that the set of PSF derived from this experiment is not 
consistent with new XRD data with much better 
statistical accuracy. A new set of PSF is derived which corrects this 
deficiency. The first diffraction peak is not dominated by the 
cation-cation-PSF and the set of PSF shows a remarkable similarity to
the PSF of glassy GeO$_2$ and SiO$_2$ determined recently. The Levesque, 
Weis, Reatto potential inversion scheme has been used to interpret the
PSF in terms of three dimensional structures. The thus derived angular 
correlations between near neighbor atoms show similarities with CuBr 
melt, reflecting the fact that both melts are based on tetrahedral 
structural units but also characteristic differences in the $+-+$ (ZnClZn)
angle distribution, the angle joining adjacent tetrahedra. Further, a 
plausibility for the enhanced electrical conductivity of ZnCl$_2$ at higher 
pressures is given, based on simulations with the equilibrated potential. 
\end{abstract}
\vspace{1cm}
\noindent\par
\end{center}

\section{Introduction}

ZnCl$_2$ and other strongly structured divalent and trivalent chloride 
melts have been intensely studied in the past \cite{Rov86,McG90,Tos93,
Akd98,Tat00}. In ZnCl$_2$ 
the tendency to form long ranged structures without periodicity is so
pronounced, that it can be easily supercooled into a glass \cite{desa}. 
It has been found \cite{Big} \footnote{This reference is called I 
in the following}, 
that the ZnZn and the ClCl next neighbor shell occur at the same distance. 
Such structures
are especially difficult to describe by ionic models of the liquid 
structure. The reason for this difficulty is asymmetry between cation and 
anion charge. If such an ionic potential is applied, the structure tend to 
be fluorite-like, i. e. the doubly charged ions further apart than the 
singly charged ions \cite{Mad93}. The inability of ionic models to 
reproduce the experimental structure has been taken as indication \cite{desa},
that the bonding in ZnCl$_2$ has a more covalent character.

In the early eighties the partial pair distribution functions of ZnCl$_2$
have been determined by Biggin and Enderby (I) with neutron diffraction and
the isotopic substitution (NDIS) technique. Wilson and Madden, when forwarding
their view that ZnCl$_2$ can be described entirely by a polarizable ion
model as opposed to a largely covalent description \cite{Mad94,Mad98} refer
to this now almost twenty years old NDIS study to evaluate the success of their
model to describe the ZnCl$_2$ structure. A special accent has been given
to the attribution of the first peak in momentum space to the ZnZn partial
structure factor (PSF). However, the ZnZn-PSF, as it has a low
weight in a total neutron structure factor, is the one which is most 
difficult to determine. It will be shown in this text, that the PSF 
determined by in (I)
are not compatible with more recent neutron \cite{How} and electromagnetic 
radiation scattering experiments \cite{Neu} and an alternative set of PSF 
is determined which corrects for these deficiencies. The potential inversion 
scheme of Levesque, Weiss and Reatto \cite{LWR} (LWR-scheme) 
is then applied to 
interpret the PSF in terms of three dimensional structures.

\section{Theoretical Summary}

The differential cross section of a liquid in a neutron 
or electromagnetic 
radiation scattering experiment can be expressed in terms of a total 
scattering function:
\begin{eqnarray}
S^{(n)}(Q)&=&\frac{\left(\frac{d\sigma}{d\Omega}\right)^{(n)}-
\sum_{i}^{N_{uc}} \nu_i b_i^2}{(\sum_{i}^{N_{uc}} \nu_i b_i)^2}+1 
\label{eq-sq-n}\\
S^{(x)}(Q)=i(Q)+1&=&\frac{\left(\frac{d\sigma}{d\Omega}\right)^{(x)}/
\sigma_{el}-\sum_{i}^{N_{uc}} \nu_i f_i^2}{(\sum_{i}^{N_{uc}} \nu_i f_i)^2}
+1\label{eq-sq-x}
\end{eqnarray}
where $\left(\frac{d\sigma}{d\Omega}\right)$ is the coherent 
differential cross section, $Q=4\pi/\lambda \sin(\theta)$, $\lambda$ the 
wavelength of the radiation and $\theta$ the diffraction angle, 
$b_i$ the coherent scattering lengths \cite{Koe91}, $f_i$ the X-ray form 
factors in the independent atom approximation \cite{Hub75}, 
$\sigma_{el}$ the scattering cross section of the free electron, $\nu_i$ the 
stoichiometric coefficient of the atom $i$, and where the sums are extending 
over the number of distinct atoms $N_{uc}$ in the unit of composition, 
ZnCl$_2$. 
These total structure functions are composed of partial structure factors
PSF $s_{ij}$:
\begin{equation}
S^{(n/x)}= \sum_{ij} w_{ij}(Q) s_{ij}(Q) \label{eq-PSF}
 \end{equation} 
with
\begin{equation}
w_{ij}(Q) = \frac{\nu_i \nu_j f_i(Q) f_j(Q)}{(\sum \nu_i f_i(Q)^2}
\label{eq-wij}
\end{equation} 
where for the neutron case $f(Q)$ has to be replaced by $b$ and the 
weighting factors  become independent of $Q$.
The partial structure factors are related to the partial pair distribution 
functions (PPDF) $g_{ij}$ via Fourier sine transformation:
\begin{equation}
r \cdot (g_{ij}-1) = \frac{1}{2\pi^2\rho_{uc}} 
\int Q \cdot (s_{ij}-1)  
\sin(Qr) dQ \label{eq-ft-n}
\end{equation}
with $\rho_{uc}$ the density per unit of composition. The PPDF $g_{ij}$
describe the probability to find an atom of type $j$ at distance $r$ from an 
atom of type $i$ relative to the bulk density.

An alternative definition of the neutron total structure factor has been 
used by Biggin and Enderby:
\begin{equation}
F(Q)=c_a^2 b_a^2 [S_{aa}(Q)-1] + 2 c_a c_b b_a b_b [S_{ab}(Q)-1] + c_b^2 b_b^2 [S_{bb}(Q)-1] \label{eq-f}
\end{equation}
where $c_a$, $c_b$ are the concentration of the atomic species, 
the $S_{aa}$, $S_{bb}$, $S_{bb}$ are partial structure factors differing from
the $s_{ij}$ in Eq. \ref{eq-PSF} by a factor of $\sum \nu$. It is noted, 
that in this definition the total neutron structure factor has the 
dimension of a cross section, while $S^{(n)}(Q)$ in Eq. \ref{eq-sq-n} it  
is dimensionless.

In order to generate three dimensional structures from the pair distribution 
functions, the potential inversion scheme of Levesque, 
Weiss and Reatto \cite{RLW} (LWR-scheme) has been applied. The idea of 
this method is based on the equation:
\begin{equation}
\label{eq-hd}
g(r)= \exp \left[ \frac{-v(r)}{kT}  + g(r) - 1 - c(r) + B(r,v)\right]
\end{equation}
relating the pair distribution function and the pair potential
$v(r)$, where $c(r)$ the direct correlation function and 
B(r,v) the bridge function. Starting with a first guess of the bridge function,
e. g. neglecting it completely, a first guess of the potential $v^{(1)}$ 
is calculated.  A Monte Carlo (MC) or Molecular Dynamics (MD) simulation 
gives $g(r)^{(1)}$ and $c(r)^{(1)}$ belonging to $v^{(1)}$ and 
thus $B(r,v^{(1)})$. $B(r,v^{(1)})$ is expected to be a better approximation 
for $B(r,v)$ than completely neglecting it.
Thus, substituting $B(r,v^{(n-1)})$ for $B(r,v)$ in equation \ref{eq-hd} gives 
the LWR iteration formula
\begin{eqnarray}
\nonumber &&v^{(n)}/kT=v^{(n-1)}/kT + \mathrm{ln} (g^{(n-1)}/g^{(exp)}) \\&&+ c^{(n-1)} - c^{(exp)} - g^{(n-1)} + g^{(exp)}\label{eq-iter}
\end{eqnarray}
The empirical potential Monte-Carlo (EPMC) scheme \cite{EPMC} 
is closely related to  equation \ref{eq-iter}, 
but considers the logarithmic term only. The complete form of Eq. \ref{eq-iter}
has been preferred here as this scheme shows faster  convergence \cite{RLW}. 
The applicability of the RLW scheme to polyatomic systems has been shown 
by \cite{kahl}. An example of the application of this technique has been given recently in \cite{phyb}.

\section{Deduction of the PPDF}
Biggin and Enderby (I) have determined 
the PSF and the PPDF of ZnCl$_2$ melt with 
neutron diffraction and isotopic substitution   of the
chlorine atom.  This technique uses the fact that the neutron scattering 
power of the two chlorine isotopes for neutrons are significantly different.
Thus, assuming the independence of the structure from the isotopes and the 
preparation involved, a set of three equations of the type 
given by Eq.~\ref{eq-PSF} 
are obtained. 

By coincidence the weighting factors $w_{ij}$  of the 
PSF in a neutron diffraction experiment on Zn$^{37}$Cl$_2$ and an 
electromagnetic radiation scattering experiment are the same, except for the 
small $Q$ dependence of the $^{(x)}w_{ij}$. 
Thus, $S^{(n)}_{\rm Zn^{37}Cl_2}$ and
$S^{(x)}_{\rm ZnCl_2}$ are equal. Fig. \ref{fig-comp-endi} compares these 
quantities as well as $S^{(n)}_{\rm Zn^{37}Cl_2}$ and $S^{(x)}_{\rm ZnCl_2}$
which would result from the PSF given in (I). It is evident from this figure,
that the PSF from (I) are incompatible with the X-ray result.

\begin{figure}[phtb]
\begin{center}

\rule{0pt}{0pt}
\end{center}
\caption{\label{fig-comp-endi}
Comparison the total neutron scattering function of 
Zn$^{37}$Cl$_2$ (I) and the electromagnetic radiation scattering 
function of ZnCl$_2$ \cite{Neu} \\
$F_{Zn^{37}Cl_2}(Q)=0.3956 \cdot 10^{-28} {\rm m}^2 [S^{(n)}_{Zn^{37}Cl_2}-1]$ 
(Eq. \ref{eq-f}) (open diamonds, from I) 
is compared with $0.3956 10^{-28} {\rm m}^2 i(Q)$ 
(small crosses, from \cite{Neu}) as well as  $F_{Zn^{37}Cl_2}(Q)$
(solid line) and $0.3956 10^{-28} {\rm m}^2 i(Q)$ (broken line), 
which would result from the PSF given in I. }
\end{figure}
Wilson and Madden determined the PSF of the ZnCl$_2$ system 
from a computer simulation using a polarizable ion model.
In Fig. \ref{fig-comp-mad} the results of a similar comparison of the 
experimental electromagnetic radiation scattering function $i(Q)$ with the
result obtained for the simulated polarizable ion PSF via Eq. \ref{eq-PSF} and
 \ref{eq-wij} are shown. Again, the 
incompatibility of the polarizable ion PSF with the electromagnetic radiation 
scattering experiment can be seen. The overestimation of the
low-Q peaks is noted.

\begin{figure}[!htb]
\begin{center}

\rule{0pt}{0pt}
\end{center}
\caption{\label{fig-comp-mad} Comparison of the experimental 
total X-ray structure factor $i(Q)$ (solid line) with $i(Q)$ calculated 
from the PSF resulting from the simulation of Wilson and Madden 
\cite{Mad94} (symbols)}
\end{figure}

The structure of ZnCl$_2$, as a two component system, is described 
by three PPDF/ PSF. A set of three independent 
equations~\ref{eq-sq-x} is hence necessary to determine the PSF.
The first two are the hard X-ray $i(Q)$\cite{Neu} and
the neutron $S(Q)$ of
of ZnCl$_2$ in natural isotopic composition \cite{How}, the procedure 
to obtain the third independent information is described in the following.
The ZnCl-PSF is dominated largely by the 
contributions of the first peak in real space at 2.3 \AA 
(cf. Fig.~\ref{fig-zncl-ppdf}). 
This peak is separated from the rest of the structure in real space
even in the total $g^{(x)}(r)$\cite{Neu}. The ZnCl-PSF of (I) was, hence, 
split by Fourier-filtering into the part 
originating from the first peak in real space and the 
remaining contribution:
$^{(n)}S_{\rm ZnCl} = S^{(n,FP)} + S^ {(n,rem)}$ 
and $S^{(n,FP)}$ was removed from the ZnCl-PSF
$S^{(n,rem)}$ shows oscillations exceeding the error level 
only for $Q<4$\AA$^{-1}$ and is set to 
zero for $Q>4$\,\AA$^{-1}$. $s^{(FP)}$ has been determined from 
the hard X-ray total structure factor 
$i(Q)$. A new $s_{\rm ZnCl}$ is then given by 
the contribution of the first peak from the hard X-rays and 
neutron derived $s^{(n,rem)}$:
$s_{\rm ZnCl} = s^{(x,FP)} + s^{(n,rem)}$. This gives the missing piece 
of information to determine all three PSF/PPDF.

Relying on one partial structure factor
from the NDIS experiment more than on the other may seem somewhat 
arbitrary, but two reasons justify the choice: First, the $s_{\rm ZnCl}$
is less affected by statistical errors then $s_{\rm ZnZn}$ 
since its weighting factor is large and, second, $s^{(FP)}_{\rm ZnCl}$ 
can be taken from the hard X-ray data. 

The resulting partial structure 
factors are shown in Fig.~\ref{fig-zncl-parti}. It is noted, that
the attribution of the first peak in $Q$ space to $s_{\rm ZnZn}$, 
which seemed obvious from the old NDIS data is not reproduced here.
 Instead, the maximum at 1\AA$^{-1}$ is mainly a result of the sine wave in
$s_{ZnCl}$ belonging to the first peak in $r$ space. It is underlined, 
that the PSF in Fig.~\ref{fig-zncl-parti} are consistent with all available
diffraction information and that a tree dimensional arrangement of atoms 
exist, which is consistent with these PSF (this is shown later in this paper).
It is noted that the PSF of vitreous germanium-dioxide determined 
by Price {\it et al} 
\cite{Pri98} and the PSF of vitreous SiO$_2$ calculated from the ab-initio
simulation by Sarntheim {\it et al} \cite{Sar95} 
show a remarkably similar behavior. It has been hypothesized 
\cite{Ell91}, that the first diffraction peak in 
AX$_2$ type glass formers is generally dominated by the cation-cation PSF.
Together with the ZnCl$_2$ melt, there are now three prominent glass formers
of AX$_2$ composition, where no indication for such a domination is found.
\begin{figure}[htb]
\begin{center}

\rule{0pt}{0pt}
\end{center}
\caption{\label{fig-zncl-parti}Comparison of the simulated structure 
factor with the experiment at T=873K\\
Left: Comparison of the experimental (cf. preceding section)
partial structure factors (solid line)
with the simulation (broken line). \\
Right: Comparison of the experimental 
total structure functions $i(Q)$ und $F(Q)$ (data points), the recomposited 
experimental partial structure functions (solid line) and the simulated 
PSF (broken line). Solid and broken line are nearly 
indistinguishable.}
\end{figure} 
\begin{figure}[ht]
\begin{center}

\rule{0pt}{0pt}
\end{center}
\caption{\label{fig-zncl-ppdf}Partial pair distribution function of ZnCl$_2$
from the experiment (Solid line) and the simulation (broken line).}
\end{figure}

\section{Simulation results}

In Fig.~\ref{fig-zncl-parti} and \ref{fig-zncl-ppdf} the PSF and the 
PPDF deduced as described  in the 
preceding section and the result of a MC simulation with the converged 
potentials from the LWR scheme is shown. 
Further, a comparison to the total neutron and X-ray structure
functions is given (cf. Fig.~\ref{fig-comp-mad}).
The total structure functions agrees with the simulation results
almost perfectly except at the lowest $Q$-values. A 
slight overestimation of the periodicity in $g_{ZnCl}$ and 
$g_{ClCl}$ with respect to the experimental functions is to be 
remarked, transforming into differences of the height of the
peak at 2\AA$^{-1}$ in $Q$ space. 
It is, however, possible that 
this is due to the imperfection of the experimental functions composed
from quite different sources and the simulation correct for 
these imperfections to some extend, due to the 
restrictions imposed by the necessary space filling of a simulation box of 
given size. A snapshot of the simulated ZnCl$_2$ melt is shown in 
Fig.~\ref{zncl2bild}.

\begin{figure}
\caption{Snapshot of simulated ZnCl$_2$ melt\label{zncl2bild}.\\
Dark spheres are
Zn-Atoms and light spheres Cl-Atoms, adjacent atoms are connected with solid 
cylinders. Shown is neighborhood of the Zn-atom in the middle of figure up to 
the forth shell.}
\end{figure}
 
In Fig.~\ref{fig-zncl2-ang} the angular correlations between 
neighboring atoms
resulting from the MC-simulation with the converged LWR potential 
are shown. A distance criterion has been applied to decide, 
whether atoms are neighboring, ''bonded'', or not. 
The cut-off radius for both ZnZn and ClCl pairs, 4.8\AA,  corresponds 
roughly to the first minima in the partial pair distribution 
functions. For the ZnCl PPDF the minimum between first and second 
maximum is very pronounced, and the choice of the cut-off radius, 2.8\AA, 
is evident. 
The $-+-$ (ClZnCl) distribution, peaking at $\cos(\theta)= -1/3$, 
correspond to the geometry of the ZnCl$_4$ tetrahedra, while the 
$---$ distribution shows a pattern
characteristic for random dense packing (cf. Finney's model \cite{finn} 
for ZnCl$_2$ glass). 
These two distribution functions are very similar to the corresponding 
$-+-$ and $---$ of CuBr(l) 
determined by Pusztai and McGreevy \cite{MCG} with Reverse 
Monte-Carlo (RMC) simulation. This 
reflects the fact that this liquid also shows a 
tetrahedral ordering - Cu(I) is isoelectronic to Zn(II). On the other hand, 
there is a marked difference in the $+-+$ distribution function, the 
angle connecting adjacent tetrahedra, which is much more pronounced in the  
ZnCl$_2$ case. 
This makes plausible why a continuous random network is a 
meaningful description of the ZnCl$_2$\cite{Neu}, but not the CuBr structure. 
 Such a preference for certain $+-+$ angles restrict the possibilities to 
arrange the elemental tetrahedral units in a periodic lattice and 
makes it plausible, why the ZnCl$_2$ but not CuBr melt can be easily 
supercooled into a glassy state.
\begin{figure}[htbp] 
\begin{center}
\rule{0pt}{0pt}
\end{center}
\caption{\label{fig-zncl2-ang} Angle distribution functions of adjacent
atoms\\
}
\end{figure}

The pair potential being determined at one particular state point, it can be 
used in an NVT-Monte Carlo simulation to predict of the structure
at state points which are not easily accessible to diffraction 
experiments, assuming the state independence of that effective 
pair potential.
In  Fig.~\ref{fig-predic-pre} the changes in the pair 
distribution functions are shown, which are obtained from
a Monte-Carlo simulation using the LWR potentials 
determined above, when increasing the density by 3.3\%
6.6\% or 10\%. A ten percent density increase corresponds to an approximate 
pressure of 3000 bar (calculated from the compressibility 
at atmospheric pressure). 
\begin{figure}
\caption{\label{fig-predic-pre}
Isothermal compression of ZnCl$_2$ melt at 873 K.
\\ Left: Electromagnetic radiation total structure factor $i(Q)$ at 
atmospheric pressure (solid line), 3.3\%, 6.6\% and 10\% increased density  
(dotted,dashed dotted, dashed line)\\
Right top: ZnCl-PPDF at atmospheric pressure (triangles),3.3\% (crosses), 6.6\%
(stars) and 10\% (diamonds) increased density. 
\\
Right bottom: Change in the ZnCl-PPDF upon 10\% compression}
\end{figure}
The expected change in the
X-ray total scattering function $i(Q)$ is shown in 
Fig.~\ref{fig-predic-pre}. 
These are obviously small, the most important change is the decrease
in intensity at low-$Q$ compatible with a decrease in 
compressibility to be expected at higher pressures.
The ZnCl coordination number has a slightly increasing tendency raising 
from 3.6 to 3.7 at 10\% compression.

There has been recent interest to understand the behavior of the
electric conductivity in ZnCl$_2$ melt and strongly structured molten salts 
in general \cite{Tat00, Trulla, Aic90}. Contrary to the usual 
trend, in ZnCl$_2$ melt the conductivity increases with increasing 
pressure \cite{Aic90}.
A Monte-Carlo simulation does  
not allow, of course, the calculation of dynamic correlation functions 
like the velocity correlation function directly related to the 
conductivity.
However, it is assumed \cite{Tat00}
that a velocity determining step in the ionic conduction
in ZnCl$_2$ is the detachment of the chlorine ion from the network.
While the changes in the pair distribution in 
functions are small, the depth
of the minimum between the first and second ZnCl coordination 
shell decreases. It is probable, that small number of 
atoms at intermediate distance between the first and
second neighbor shell play a crucial role for the conductivity
and correspond to detached atom in interstitial sites
(A similar argument can be found in \cite{Kee97,Kee99}). 
Thus a plausible explanation of larger conductivity is
that these interstitial sites become accessible for more
ions and thus the number of ions accessible to charge transport 
increases.

\section{Conclusion}

Using the results of modern diffraction experiments  
improved partial structure factors and pair 
distribution functions for ZnCl$_2$ melt are obtained. The first peak 
in $Q$-space at 
1\AA$^{-1}$ has been traditionally attributed to the ZnZn partial 
structure factor. This assumption is not in agreement with the hard X-ray
diffraction experiment. A set of PPDF consistent with all 
available diffraction experiments is presented here, 
which suggests that this peak is mainly due to the ZnCl 
partial. It is also proved in this work, that this set of PSF can be mapped 
to a physical arrangement of atoms of the experimental density. 
It is shown, that the simulation result of 
Wilson and Madden, although compatible with the
old NDIS data of Biggin {\it et al} within its 
large statistical errors, shows significant deviations from 
modern hard X-ray data.

In order to interpret the partial pair distribution 
function it terms of a three dimensional structure the
Levesque, Weis, Reatto scheme has been used to deduce an empirical 
potential for this system. MC simulations with this potential are used to 
derive the angular correlations between neighboring atoms. 
It is interesting to compare the angle distribution functions of the 
ZnCl$_2$ melt, which can be easily supercooled into a glassy state
to those of the melt of the fast ion conductor CuBr.
The structure of both systems is characterized by tetrahedral structural 
units, translating in a similar $-+-$ distribution. On the other hand, the 
$+-+$ distribution , the angle joining adjacent tetrahedra, is peaked
in ZnCl$_2$ and broad in CuBr, explaining why a continuous random network is 
useful characterization of only the ZnCl$_2$ melt.

Besides of the capability to produce three dimensional
models of the ZnCl$_2$ in excellent agreement with the 
available diffraction information, the potentials 
obtained by the LWR scheme can be used to predict the
structure at state points difficult to access by diffraction 
experiments. Thus, the structure of the melt at 
elevated pressures has been obtained. It is found, that
under these conditions, the ZnCl coordination number 
increases slightly, while the overall structural changes are small. 
The increasing ionic conductivity of ZnCl$_2$ melt at higher pressures
can be explained by  the increased number of ions capable to occupy
interstitial sites in the network.

\section{Acknowledgment}
This work was supported by the Deutsche 
Forschungsgemeinschaft under grant No. Ne-584/1-2.

\rule{0pt}{0pt}


\begin{thebibliography}{99}
\bibitem{Rov86} Rovere, M., Tosi, M.~P., 1986, Rep. Prog. Phys., {\bf 49} 1001
\bibitem{McG90} McGreevy, R.~L., Pusztai, L., 1990, Proc. R. Soc. London Ser. A, {\bf 430} 241
\bibitem{Tos93} Tosi, M. P., Price, D.~L., Saboungi, M.-L., 1993, 
Annu. Rev. Phys. Chem., {\bf 44} 173 
\bibitem{Akd98} Akdeniz, Z., Price, D.~L., Saboungi, M.-L., Tosi, M. P., 1998,
Plasmas Ions, {\bf 1} 3
\bibitem{Tat00}  Tatlipinar, H., Amoruso,M.,  Tosi, M.~P., 2000, Physica B
{\bf 275} 281 
\bibitem{desa}  Desa, J.~A.~E., Wright, A.~C.,   Wong, J., Sinclair, R.~N., 
1982, J. Non-Cryst. Sol. {\bf 51}, 57
\bibitem{Big}   Biggin, S., Enderby, J.~E.  1981, J. Phys. C {\bf 14}, 
3129
\bibitem{Mad93} Wilson, M., Madden, P.~A. ,1993 J. Phys. Condens. Matter 
{\bf 5} 6833
\bibitem{Mad94} Wilson, M., Madden, P.~A. , 1994, Phys. Rev. Lett. {\bf 72} 
3033
\bibitem{Mad98}Wilson, M., Madden, P.~A. , 1998, Phys. Rev. Lett., {\bf 80} 
532
\bibitem{How}   Allen, D.~A., Howe, R.~A., Wood, N.~D.,  Howells, W.~S. , 1991,
J. Chem. Phys. {\bf 94}, 5071
\bibitem{Neu} Neuefeind, J., T\"{o}dheide, K., Lemke, A.,
Bertagnolli, H., 1998, J. Non-Cryst. Sol. {\bf 224}, 205
\bibitem{LWR} Levesque, D., Weis, J.~J., Reatto, L. 1985, Phys. Rev. Lett.
{\bf 54}  451
\bibitem{Koe91}  K\"oster, L., Rauch, H., Seymann, E., 1991, 
At. Data Nucl. Data Tab. {\bf 49}  65
\bibitem{Hub75}  Hubell, J. H. , Veigele, W.~J., Briggs, E.~A., Brown,R.~T.,
Cromer, D.~T. ,Howerton, R.~J. , 1975, J. Phys. Chem. Ref. Data {\bf 4}  471 
\bibitem{EPMC} Soper, A.~K., 1996, Chem. Phys. {\bf 202}, 295
\bibitem{RLW} Reatto, L., Levesque, D., Weis, J.~J., 1986,
Phys. Rev. A, {\bf 33} 3451
\bibitem{kahl} Kahl, G., Kristufek, K., 1994 Phys. Rev. E,  {\bf 49}, 3565 
\bibitem{phyb} Neuefeind, J., Fischer, H.~E., Schr\"{o}er, W., 2000,
Physica B,  {\bf  276-278} 481
\bibitem{Pri98} Price, D.~L., Saboungi, M.-L., Barnes, A.~C., 1998, Phys. Rev.
Lett., {\bf 81} 3207
\bibitem{Sar95} Sarnthein, J., Pasquarello, A., Car, R., 1995, 
Phys. Rev. Lett., {\bf 74} 4682
\bibitem{Ell91} Elliot, S.~R., 1991, Phys. Rev. Lett., {\bf 67} 711 
\bibitem{finn} Finney, J.~L., 1970, Proc. Roy. Soc. (London), {\bf A319} 479
\bibitem{MCG}McGreevy, R.~L., Pusztai, L., 1998, J. Phys. Condens. Matter {\bf 10} 525
\bibitem{Trulla} Trull\`as, J., Padr\'o, J.~A., 1997, Phys. Rev. B {\bf 55} 
12210
\bibitem{Aic90} Aich, R., Ismail, K.  T\"odheide, K., 1990, 
High Pressure Res. {\bf 3-5}, 607
\bibitem{Kee97} Keen, D.~A., 1997, Phase Transitions {\bf 61} 109
\bibitem{Kee99} Keen, D.~A., Dove, M. T., 1999, J. Phys. Condens. Matter 
{\bf 11} 9263

\end{thebibliography}
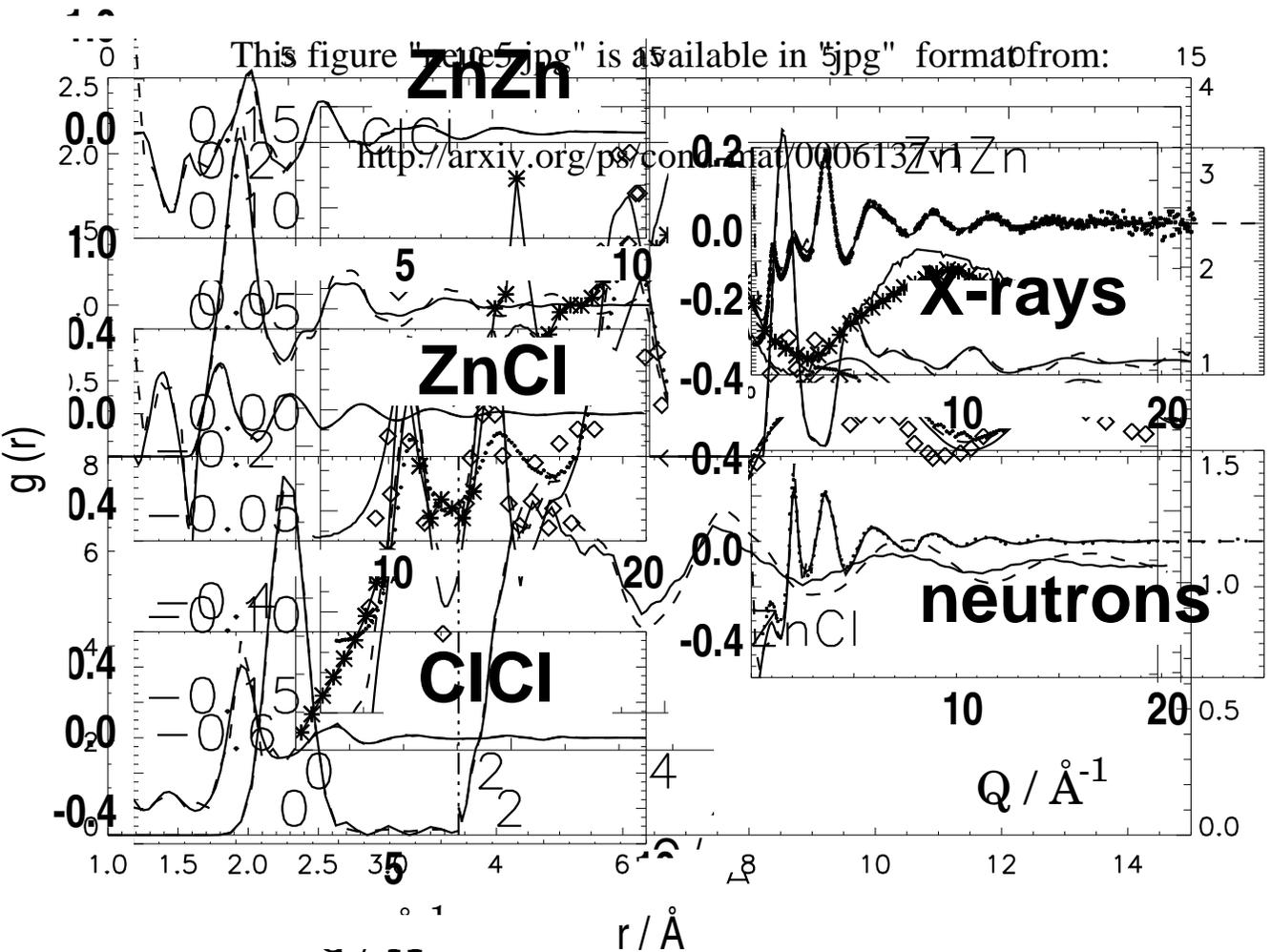
\end{document}